\documentclass[osajnl,showpacs,twocolumn]{revtex4}
\usepackage{graphicx}% Include figure files

\begin{document}

\title{Far-field spectral characterization of conical emission and filamentation in Kerr media}

\author{Daniele Faccio}
 \email{daniele.faccio@uninsubria.it}
\author{Paolo Di Trapani}
\author{Stefano Minardi$^{**}$}
\author{Alberto Bramati$^{***}$}
\affiliation{INFM and Department of Physics \& Mathematics, University of
Insubria, Via Valleggio 11, I-22100 Como, Italy}
\author{Francesca Bragheri}
\author{Carlo Liberale}
\author{Vittorio Degiorgio}
\affiliation{INFM and Department of Electronics, University of Pavia, Via Ferrata 1, I-27100 Pavia, Italy}

%\affiliation{Laboratoire Kastler Brossel, Universit\'e P. et M. Curie, Case74, 4, place Jussieu, F75252 %Paris, France}
%\affiliation{Institut de Ci\`encies Fot\`oniques, c/Jordi Girona 29, NEXUS II, E-08034 Barcelona, Spain}
\author{Audrius Dubietis}
\author{Aidas Matijosius}
%\author{Rimtautas Piskarskas}
%\author{Arunas Varanavi\v{c}ius}
\affiliation{Department of Quantum Electronics, Vilnius University, Sauletekio Ave. 9, bldg. 3, LT-2040, Vilnius, Lithuania}

\date{\today}

\begin{abstract}
By use of an imaging spectrometer we map the far-field ($\theta-\lambda$) spectra of 200 fs optical pulses that have undergone beam collapse and filamentation in a Kerr medium. By studying the evolution of the spectra with increasing input power and using a model based on stationary linear asymptotic wave modes, we are able to trace a consistent model of optical beam collapse high-lighting the interplay between conical emission, multiple pulse splitting and other effects such as spatial chirp.
\end{abstract}


%\keywords{tolerances, second order nonlinearity, photonic crystal}

\maketitle

\section{Introduction}
Filamentation and, in general, collapse, of high power laser pulses in transparent media has attracted significant attention ever since its prediction \cite{ask:1962, talanof:1964, chiao:1964} and observation \cite{hercher:1964} and, due to the complexity of the involved phenomena, continues to be a hotly debated topic. There are many reasons for this interest ranging from application possibilities such as localizing ultra-short laser pulses over long propagation distances \cite{braun:1995}, white-light laser sources for parametric amplification \cite{wilson:1997} or spectroscopy \cite{alfano:1989, rai:2000} or the formal analogy with the equations that describe nonlinear wave collapse in other systems, such as Bose-Einstein condensates \cite{noz:1990}, that are experimentally less accessible.  Numerical investigation of optical wave collapse is usually carried out starting from the nonlinear Schr\"odinger equation (NLSE) that describes the spatial evolution of a beam considering diffraction and a self-focusing term that originates from the real part of the third-order medium Kerr nonlinearity ($n_{2}$). Such an equation predicts the formation of an unstable 2D stationary solution, the so called Townes profile \cite{chiao:1964}, that, if perturbed, will either diffract or undergo catastrophic collapse. However, ultra-short laser pulses do not, in general, follow this behavior \cite{strickland:1994}: the collapse is arrested by other effects such as pulse lengthening due to normal group-velocity dispersion (GVD) or plasma defocusing and an apparently stationary propagation regime (filament) is reached. Thus the NLSE may be accordingly modified to account for space-time coupling and also for plasma generation and defocusing. These modified equations have proved to be able to describe many of the phenomena associated with pulse filamentation such as conical emission (CE) \cite{luther:1994}, super-continuum (SC) generation, pulse steepening \cite{gaeta:2000} and splitting \cite{ranka:1998}. One of the main features that is emerging is the importance of space-time coupling.
Indeed, there exist regimes characterized simultaneously by ultra-tight focusing and ultra-short pulse length in which the nonlinearity couples the spatial and temporal dynamics following an underlying geometry dictated by the modulational instability (MI) gain profile \cite{luther:1994,trillo:2002}. In such cases it is preferable to avoid space-time separation and refer directly to the geometrical properties of the space-time environment. A measure of the importance of these new concepts is given, for example, by the discovery of nonlinear X-waves \cite{valiulis:2001, ditrapani:2003, conti:2003, ottavia:2003}, chaotic spatio-temporal fragmentation due to space-time MI \cite{salerno:2004}, red solitons \cite{minardi:2003}, X-waves in pulse filamentation \cite{kolesik:arxiv2003} and stationary conical waves supported by nonlinear losses \cite{dub:2004}. The complexity of these issues requires a careful examination of the experimental methods employed, as space-time coupled phenomena should be investigated with adequate instruments.\\
 We note that the major part of laser physics diagnostics is based on the separation of spatial and temporal effects leading, for example, to the widely used concepts of carrier spatial and temporal frequencies, beam walk-off and group velocity, diffraction and dispersion etc., i.e. either purely spatial or purely temporal quantities. This conceptual space-time division is also reflected in the standard experimental characterization methods also adopted for the study of beam filamentation. We have on the one hand, near and far field imaging that give information on the spatial profiles but ignore the temporal profile. On the other hand, temporal auto-correlation traces or frequency resolved optical gating (FROG) give a precise measurement of the temporal profile at the medium output of the whole (or a particular portion of the) filament \cite{ranka:1998, diddams:1998, bernstein:2003}, thus losing any information regarding the (transverse) space-dependent temporal profile. The limitations of FROG measurements have been partly overcome with the so called SPIDER technique \cite{gallmann:2001} that gives the temporal profile across one spatial dimension \cite{dorrer2:2002}
and by use of a particular 3D optical gating method \cite{potenza:2004, trull:2004} that provides the full spatio-temporal intensity profile of femtosecond pulses and has indeed allowed a full dimensional intensity space-time characterization of filaments in water \cite{mat:2004} and of X-waves in $\chi^{(2)}$ media \cite{trull:2004}. Although extremely powerful, this  technique requires the use of two separate, high power, synchronized laser sources, the first providing the pulse under investigation and the second a pulse that must have an appreciably shorter time duration in order to guarantee a high (temporal) resolution. Moreover, this method is based on a second-order nonlinear up-conversion process: due to the large (spatial and temporal) bandwidths involved it is necessary to resort to very thin (20 $\mu$m or less) conversion crystals which in turn lead to low output powers and the necessity to integrate over many pulses. \\
Here we propose another possible experimental investigation tool, namely far-field spectral ($\theta-\lambda$) characterization, that may give readily accessible details of the spatio-temporal pulse profile in single shot acquisition.  This method can be used as a general investigation tool for phenomena that involve space-time coupling and we show its application to pulse filamentation in normally dispersive Kerr media. To the best of our knowledge there are only a few papers in literature that show the angular ($\theta-\lambda$) spectra of Kerr-induced filaments\cite{strickland:1994, mikalauskas:2002, dub:2003} without actually paying attention to the details these may contain and certainly no systematic study in this sense has been carried out before. The data are tentatively explained with the aid of a simple interpretation model, based on a linear description of the stationary asymptotic light wave modes that act as attractors during the pulse evolution \cite{modos}.\\

\section{Survey of beam collapse and filamentation of fs pulses}
In this work we have used a normally dispersive Kerr medium with a fixed total length and the only variable parameter was the input pulse energy. Therefore, we give here a brief overview of the expected pulse evolution for increasing input energies based on the existing literature.\\
\begin{figure}
		\includegraphics[width=7cm]{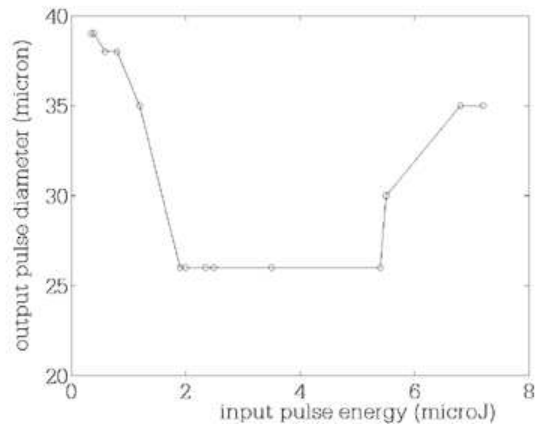}
	\caption{\label{fig:dVSE} Output beam diameter for varying input pulse energy. All measurements reported in this work were performed in the energy interval 2-5.6 $\mu$J where single filamentation was observed.}
\end{figure}
For very low powers no self focusing (SF) occurs and the beam behaves linearly. However, with increasing power the threshold for catastrophic self-focusing  is reached and the beam will start to spatially contract. A simple expression for the critical power for cw gaussian beams is given by \cite{marburger:1975}\\
\begin{equation}
\label{eq:pcrit}
P_{crit}=\frac{3.77 \lambda^{2}}{8 \pi n_{0} n_{2}}
\end{equation}
where the nonlinear refractive index is given as a function of the pulse intensity $I$ by $n=n_{0}+n_{2}I$. If the medium dispersion is also accounted for then, in general, the threshold for SF will be higher for short pulses than for longer ones due to the damping effect of normal GVD and eq. \ref{eq:pcrit} should be accordingly modified \cite{chernev:1992, luther2:1994}. It has also been noted that the threshold for SF coincides with that for spectral broadening \cite{brodeur:1999}, another manifestation of the material nonlinearity in the form of a rapidly time varying self phase modulation (SPM). For powers just above $P_{crit}$ the beam will contract spatially and the energy will move toward the back of the pulse and will eventually lead to pulse steepening and the formation of a shock wave at the trailing edge \cite{gaeta:2000, zozulya:1999, rothen:1992}. The intensities reached by the pulse give rise to efficient CE while pulse steepening may explain the observed higher conversion efficiencies for the anti-stokes (blue-shifted) components \cite{gaeta:2000}. As the power is further increased ($P\sim 1.5-2 \cdot P_{crit}$) the pulse will undergo temporal splitting \cite{rothen:1992, zozulya:1999} due to the effect of normal GVD. Even higher powers  ($P > 3 \cdot P_{crit}$) will lead to a pulse that evolves by pushing the energy toward the leading edge \cite{zozulya:1999, mlejnek:1998, mlejnek:1999, kosareva:1997}. The peak intensities have been shown to be sufficient to excite a plasma with a large enough concentration as to even compensate the Kerr induced refractive index variation \cite{brodeur:1999, tzortzakis:2001}.  The same plasma will eventually reduce the peak intensity through absorption, but for sufficiently high powers multiple pulse reformation occurs with each new pulse rising from the trailing background power \cite{mlejnek:1998, mlejnek:1999}. Along side an evident CE it has been observed that a large part of the generated spectrum is actually emitted along the pulse propagation axis \cite{bloembergen:1977}. This has been explained by taking into account a possible Kerr-lensing effect induced by the pulse itself \cite{brodeur:1999} or possibly by the formation of an effective waveguide \cite{nibbering:1996}. From this brief overview we can see that a large number of phenomena are involved, some occurring simultaneously while the presence of others depends on the particular input conditions. In the following sections we shall describe the experimental investigation method and results that shed some new light on the details of the filamentation process.
 
\section{Experimental setup}
We present data concerning filamentation in a 15 mm long, type II, lithium triborate (LBO) crystal. Such crystals are usually used for experiments involving the second order ($\chi^{(2)}$) nonlinearity, however we rotated the crystal so that all second order processes are severely phase-mismatched (and thus negligible) whilst still maintaining the pump pulse propagation axis perpendicular to the input facet. The input laser pulse is delivered from a frequency doubled, 10 Hz Nd:glass mode-locked and regeneratively amplified system (Twinkle, Light Conversion) and has 200 fs duration and 527 nm central wavelength. Spatial filtering of the pulse before entering the sample is necessary to guarantee a uniform gaussian-like profile that will thus seed only one filament at a time (i.e. no spatial break-up into multiple filaments is observed). The beam has a FWHM of 5 mm diameter and is then focused onto the LBO crystal with a 50 cm focal length lens placed at 50.1 cm from the crystal entrance facet.\\
The angular spectra of the filament have been detected by an imaging spectrometer (Oriel Instruments, 77250-M with a 1200 l/mm grating) placed after the crystal. The device
reconstructs without distortion the entrance-slit at the output plane (the slit image is not on a curved surface, as would occur for
conventional optics), with the different frequency components at different
lateral positions.
By placing the entrance slit in the focal plane of a
focusing lens, the angular frequency distribution can be detected in a single
shot. The actual spectra are captured by placing a CCD camera in the monochromator output imaging plane: the central regions require a very high dynamical range 16 bit CDD camera (Andor EEV 40-11) in order to avoid saturation and loss of low-power details whereas the outer regions were characterized using a higher spatial-resolution 8 bit camera (Pulnix TM-6 CN). Note that only the angular distribution in the slit plane will be
monitored, which is sufficient as long as the axial symmetry of the process
is preserved. In acquiring the angular spectra it is very important to
guarantee the possibility of single shot acquisition. In fact while the
envelope spectral shape gives the information on the small-scale structure
formed in the near field, it will be the  fine interference-fringe
structure in the spectra what reveals the extended feature in the
space-time domain. For example, a single X-wave, a couple, a train and even
a chaotic gas of X waves will all produce the same envelope angular
spectrum, while the difference between the possible realizations will appear only
in the modulated fine structure of the spectrum.  These fine details will fluctuate slightly from
shot to shot, owing to input-pulse energy, duration and diameter
fluctuations. \\
\begin{figure}
		\includegraphics[width=7cm]{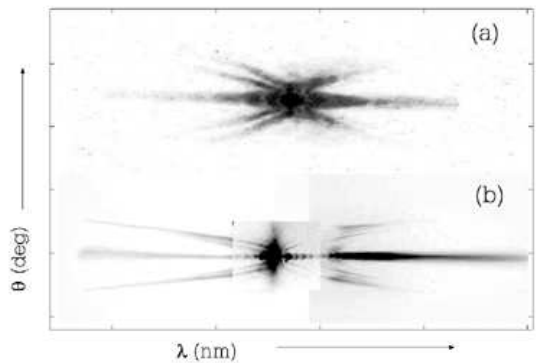}\\
	\caption{\label{fig:X} Two examples of pulse-filamentation far-field spectra taken for two different materials: (a) in 15 mm of water, $P \sim 2\cdot P_{crit}$ and (b)in 15 mm of LBO, $P \sim 1\cdot P_{crit}$. (b) was obtained from three separate images each one spanning a different wavelength range (420-510 nm, 510-540 nm, 540-650 nm)}
\end{figure}
  \begin{figure}[t]
		\includegraphics[width=7cm]{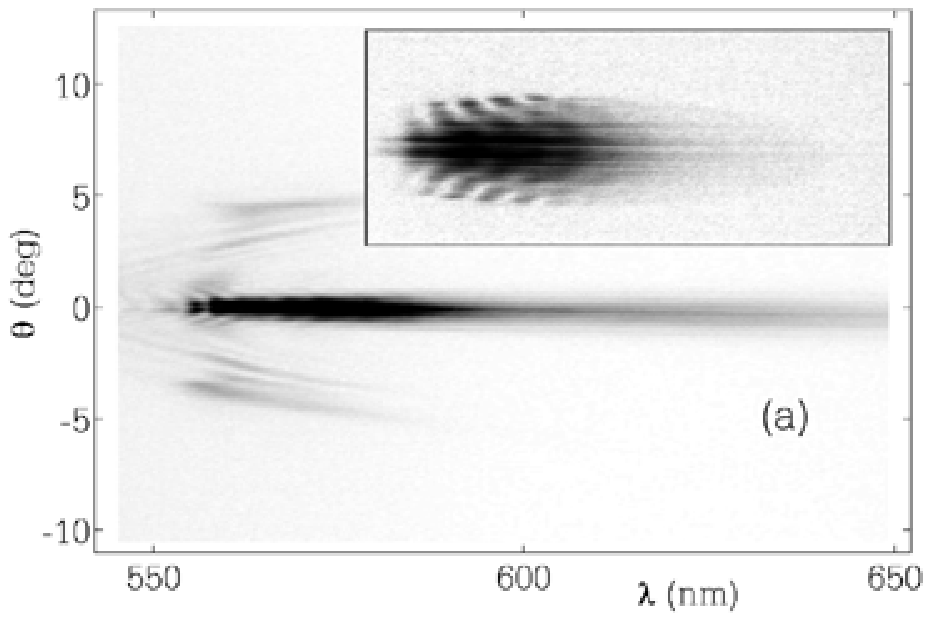}\\
		\includegraphics[width=7cm]{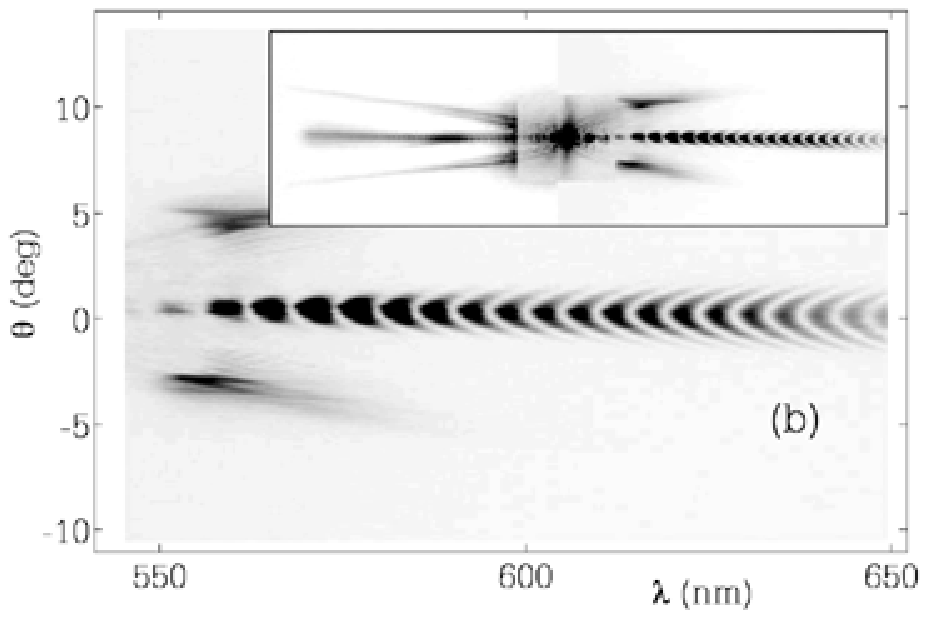}\\
		\includegraphics[width=7cm]{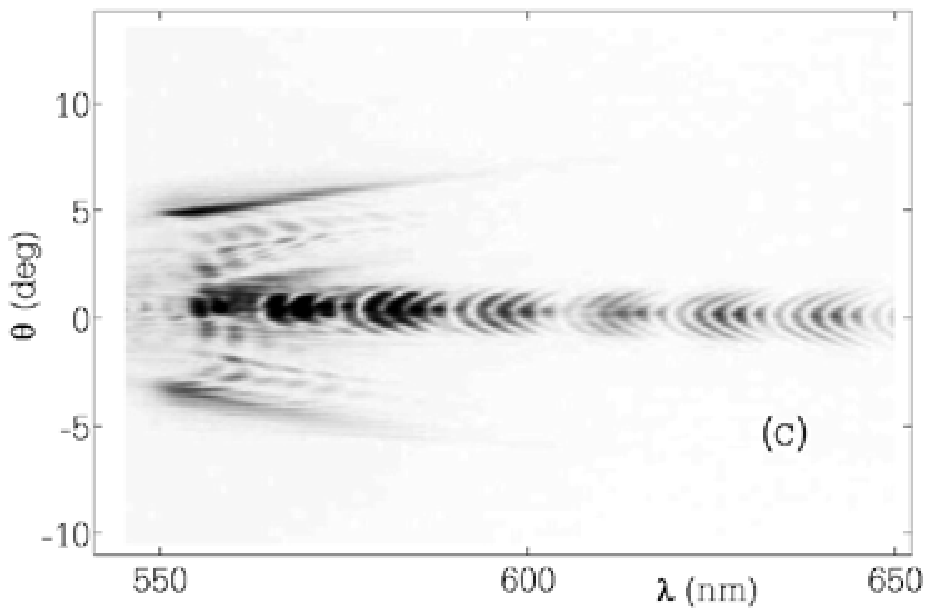}\\
	\caption{\label{fig:evol} Stokes component of the measured far-field spectra. (a) Input pulse energy $E_{in}=2$ $\mu$J: in the inset we show another spectrum taken for a slightly smaller $E_{in}$ that highlights the modulation fringes. The total spectrum is showed in Fig \ref{fig:X}. (b) $E_{in}=3$ $\mu$J: in the inset we show the total recorded spectrum. (c)  $E_{in}=4.8$ $\mu$J. Note the different modulation patterns in the three measurements.}
\end{figure}

\section{Experimental results}
In figure \ref{fig:dVSE} we show the filament $1/e^{2}$ diameter against input pulse energy. We see that as power increases the beam diameter contracts until it reaches a ``stationary'' value - we identify the lowest energy for which this occurs as the corresponding critical threshold energy, $E_{th}$. We therefore performed all our measurements in the energy range for which stable single filamentation occurs, i.e. from 2 $\mu$J to 5.6 $\mu$J. In order to illustrate that the phenomenum under inspection is a general feature of optical pulse propagation and is not particularly related to the specific material, we show in figure \ref{fig:X} two angular spectra, the first obtained in water (fig.\ref{fig:X}(a)) with $E\sim 2 \cdot E_{th}$ and the second in LBO (Fig. \ref{fig:X}(b)) with $E\sim 1 \cdot E_{th}$. Fig. \ref{fig:X}(b) was obtained from three separate images each one spanning a different wavelength range (420-510 nm, 510-540 nm, 540-650 nm) so as to minimize the effects of ``blooming'' from the central high-intensity region while still keeping significant detail in the lower intensity regions. Note also that very similar spectral features have also been reported elsewhere relative to filamentation of picosecond pulses in air \cite{mikalauskas:2002}.
Although Fig.\ref{fig:X}(a) has fewer details due the large number (30) of shots over which the profile for water was integrated, the spectra have in common a definite ``X-like'' pattern with a strong on-axis emission. We obtained similar figures also for fused silica. \\
 We shall now look more closely at the angular spectra, paying particular attention to the details. In figures \ref{fig:evol}(a), (b) and (c) we show the angular spectra for input energies $E_{in}$ equal to 2, 3 and 4.8 $\mu$J, respectively, thus mapping the evolution of the pulse structure for increasing powers. In all these figures we have focused our attention on the Stokes part of the spectrum: the complete recorded spectrum for $E_{in}=2$ $\mu$J is shown in Fig. \ref{fig:X}(b) and that for $E_{in}=3$ $\mu$J in the inset of Fig. \ref{fig:evol}(b). We first note the strong on-axis (i.e. for small transverse wave-vectors or angles) emission (AE) that extends well both into the blue Stokes and red anti-Stokes regions, characteristic of pulse filamentation and subsequent continuum generation. If considered in frequency rather than in wavelength, the anti-Stokes components have a larger extension ($\sim0.9$ fs$^{-1}$) than the Stokes components ($\sim0.7$ fs$^{-1}$), in agreement with literature \cite{brodeur:1999, gaeta:2000, liu:2003b}. Together with the low-angle emission there is also a distinct ``X'' pattern, a signature of CE with a much more pronounced extension of the anti-Stokes component ($\sim0.9$ fs$^{-1}$ to be compared with the $\sim0.2$ fs$^{-1}$ of the Stokes part). All of the recorded spectra show a surprisingly regular pattern for small $\theta$. This pattern is not so obvious in Fig. \ref{fig:evol}(a) so we have included in the inset the Stokes spectrum for a different laser shot in which it is much clearer. The fringes are centered at $\theta=0$ with parabolic-like dependence on wavelength (similar features have also been observed in air with longer input pulse durations \cite{mikalauskas:2002}). It is interesting to note how the sign of the fringe curvature inverts in passing from $E_{in}=2$ $\mu$J to $E_{in}=3$ $\mu$J and then shows a further sovra-modulation for $E_{in}=4.8$ $\mu$J (Fig. \ref{fig:evol}). We note that this behavior was found also in other materials (e.g. water) and under different focusing conditions with the only difference being the actual input energies at which the various modulation patterns are observed.\\
 \section{Discussion}
\begin{figure}[!t]
		\includegraphics[width=7cm]{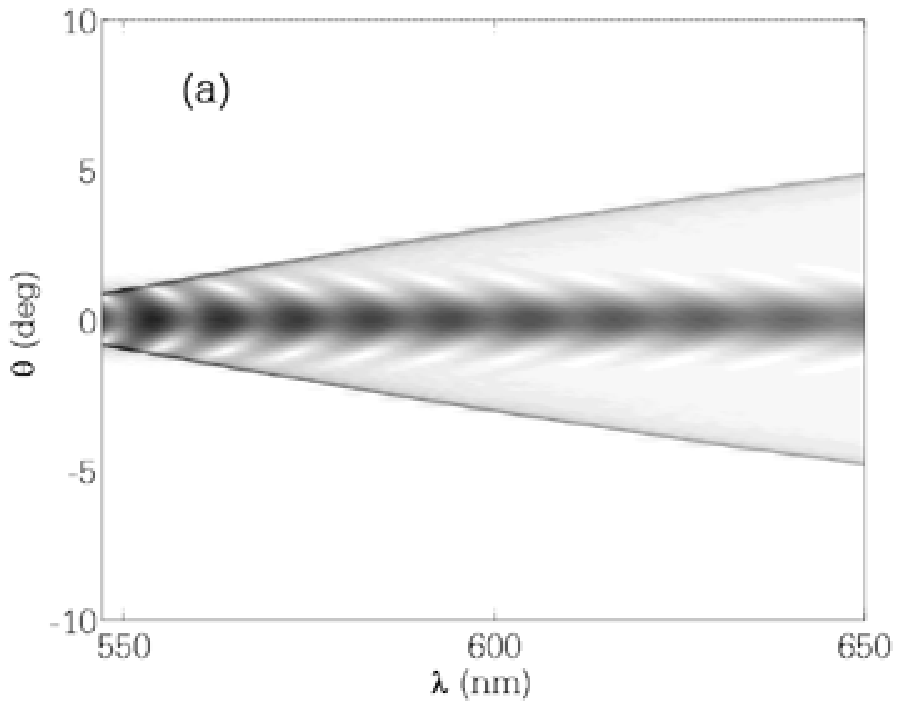}\\
		\includegraphics[width=7cm]{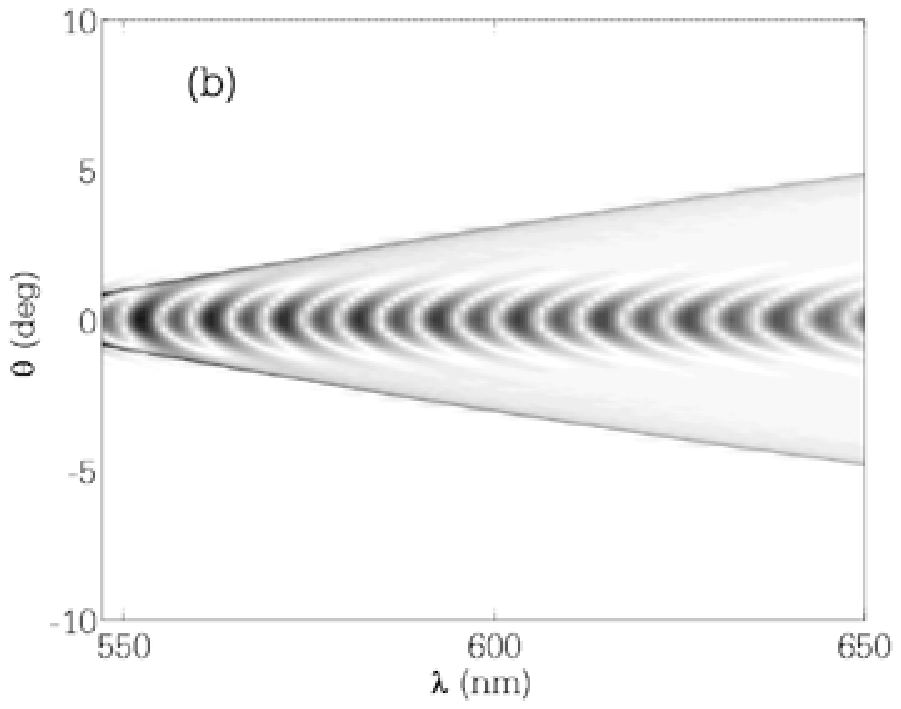}\\
		\includegraphics[width=7cm]{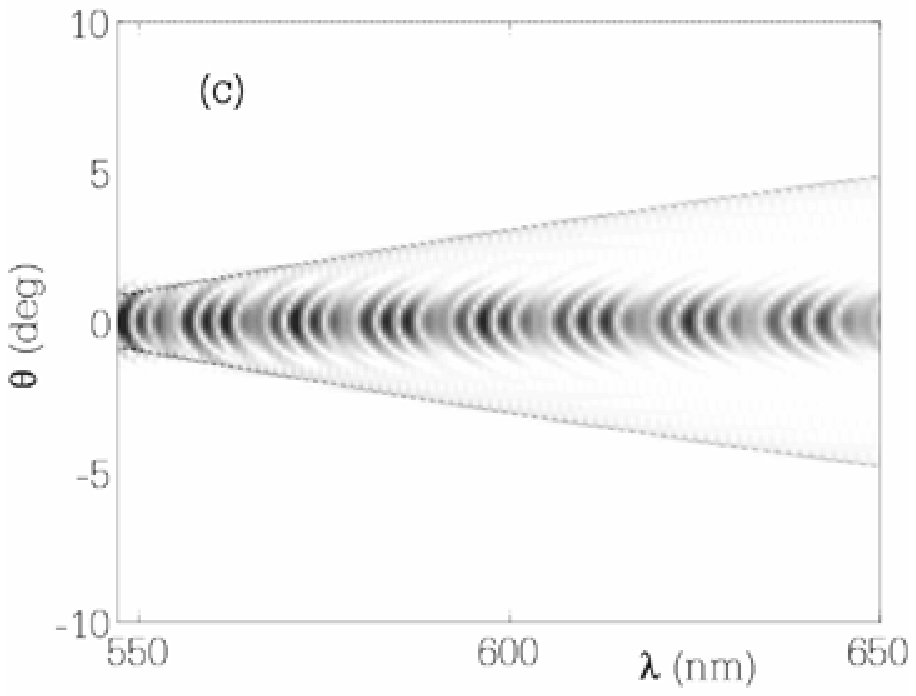}\\
	\caption{\label{fig:evol_num} Calculated far-field spectra of stationary linear states. (a) a gaussian pulse with spatial chirp $\alpha_{s}=-2$ and temporally delayed by +200 fs, with respect to an X-wave, (b) two equal-intensity spatially chirped gaussian pulses temporally shifted with respect to an X-wave by -220 fs ($\alpha_{s}=+2$) and +220 fs ($\alpha_{s}=-2$) and (c) three equal-intensity spatially chirped gaussian pulses temporally shifted by -300 ($\alpha_{s}=+2$), -200 ($\alpha_{s}=+2$) and +300 fs ($\alpha_{s}=-2$) with an X-wave in t=0. All gaussian radial diameters are 25 $\mu m$ at $1/e^{2}$.}
\end{figure}

\begin{figure}
		\includegraphics[width=7cm]{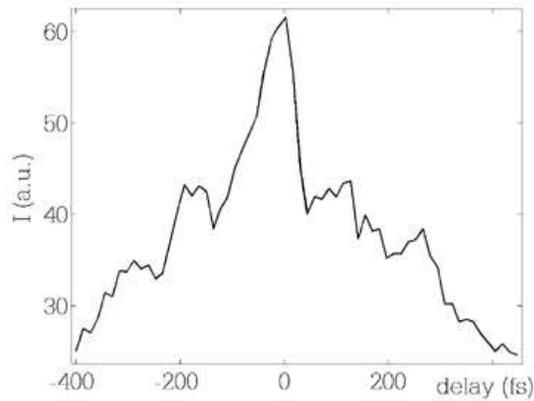}\\
	\caption{\label{fig:autocor} Autocorrelation trace of of the ouput pulse filament clearly showing that the pulse has split into three sub-pulses with a delay of 600 fs between the most external ones. $E_{in}\sim 4$ $\mu$J.}
\end{figure}

 Despite the large amount of data available in literature, to the best of our knowledge none of the numerical simulations shown to date display a combination of the main features we have measured. Namely, these are the distinct X-arms, the strong and largely extended axial emission, the periodic modulation of the axial emission and, finally, the inversion of this modulation pattern with increasing input power. Given this large difference, we try to give an explanation of the spectra by studying the properties of a linear superposition of stationary states, with which we approximate the istantaneous wave forms (into which the actual pulse shape may be decomposed) inside the material. The nature of the linear states will depend on the nature of the associated spectral shape so that CE will be related to an X-like wave \cite{trillo:2002} whereas AE may be simply related to a gaussian-like spatio-temporal profile. We underline that the following discussion is not aimed at explaining beam filamentation but rather at trying to justify the experimental spectral features starting from results presented in literature.\\
 We start by analyzing the spectrum for $E_{in}=2$ $\mu$J. The input power is just above $P_{crit}$ so that during propagation the pulse starts to collapse - in doing so a strong CE is initiated and finally the energy is moved to the trailing edge of the filament, as discussed above. Therefore we may describe the overall pulse as an X-wave (associated to the CE) and a trailing gaussian profile. The X wave is described following the recipe given elsewhere \cite{modos} with the input parameters taken as the material refractive index and dispersion relations and the total bandwidth of the actually measured CE (110 nm). The Gaussian profile is chosen so that it has a $1/e^{2}$ width equal to that measured (26 $\mu$m) and a temporal bandwidth corresponding to the measured AE bandwidth (230 nm). We note that all the results described below do not actually depend on the particular function chosen to describe the gaussian-like profile and no substantial variations were observed using a super-gaussian or hyperbolic-secant profile with respect to a simple gaussian form. Furthermore we may expect a temporal delay between the two profiles and indeed it is the interference between these two that gives rise to the measured modulation patterns. However, we are still missing an ingredient. If we take the above described pulses, and Fourier-transform the sum of these so as to obtain the angular spectra we observe only straight, vertically aligned interference fringes. In order to explain the curvature of these we must also introduce a transverse spatial chirp (i.e. a phase front curvature) into the gaussian profile. It is well known that in the presence of GVD and/or SPM the pulse may develop a strong temporal chirp. It has also been noted that in the simultaneous presence of SPM and SF, the pulse may also develop a strong frequency dependent mode size (FDMS) \cite{cundiff:1996}, and a spatial chirp. In analogy with a linear temporal chirp \cite{akhmanov} we may write the complex spatial amplitude as $A(r) = exp[-(1+j\alpha_{r})\cdot(r/\sigma)^{2}]$ were $\sigma$ is the beam width and $\alpha_{r}$ is the spatial chirp parameter. Fig. \ref{fig:evol_num}(a) shows the ($\theta-\lambda$) distribution for an X wave and a spatially chirped gaussian pulse with $\alpha_{r}=-2$ and temporally delayed by $\tau = -200$ fs, thus approximating the power build-up at the trailing edge of the pulse expected for low input powers. The absolute value of $\alpha_{r}$ was chosen so as to match the angular divergence of the AE and the sign so as to match the measured fringe curvature direction. Indeed, as can be seen the axial component shows a definite curved fringe pattern in close agreement with that shown in figure \ref{fig:evol}(a). A negative spatial chirp (i.e. a defocusing phase-front curvature) could be due to a delayed-plasma induced defocusing. However we note that another picture is also possible and that is one in which the spatial chirp has opposite sign ($\alpha_{r}=+2$) and the gaussian pulse (shock wave) is in front of the now trailing X-wave. Although a leading shock front with a focusing wave-front is in fact compatible with a strong SF regime, we believe the picture of a trialing shock wave (at input powers near threshold) to be more acceptable in the frame of the present literature.\\
 Regarding the presence of a strong axial emission, the scientific community is still lacking a widely accepted theory although the most often used explanations involve the formation of an effective waveguide \cite{nibbering:1996} induced by the balance of plasma defocusing and SF or just from the effect of SF alone \cite{brodeur:1999}. However SF is just the spatial manifestation of SPM. The temporal analogue generates new frequencies and, in the spirit of this paper, the two should not be considered separately. Indeed, it is the spatio-temporal effect of SPM that explains CE, so it is not obvious why at a certain point of the pulse evolution SPM should split into separate spatial (SF) and temporal (spectral broadening) phenomena. It could be possible that nonlinear losses (NLL) play an important role. Numerical simulations confirm that in the presence of NLL a high intensity pulse will develop a flat top (as also reported in literature \cite{polyakov:2001}) and, correspondingly, the spectral components generated via SPM will be strongly limited. This should occur in both the spatial and temporal coordinates. However there may exist mechanisms which break the spatio-temporal symmetry, such as Raman nonlinearity or plasma generation, which in turn could force the wave collapse in the sole temporal dimension, i.e. the formation of a temporal shock-wave with a spatial flat-top. In this case we would observe AE, an efficient broadening of the temporal spectrum with spatial wave-vectors aligned along the input pulse propagation axis.\\
  As for the spatial chirp, we note that the fringe pattern for higher powers is qualitatively different from that observed near $P_{crit}$. Indeed not only is the curvature inverted but the modulation is also much deeper and it is not concentrated near the central carrier frequency but rather becomes clearer at frequency shifts greater than those obtained by CE. If we take into account these features we realize that the interference is not due to an interplay between a gaussian and an X profile but rather between two (or more) gaussian-like pulses. Moreover, the chirp of the two pulses must have opposite sign and similar absolute value if such sharp and deep modulation patterns are to be explained. In Fig. \ref{fig:evol_num}(b) we kept $\alpha_{r}=-2$ for the rear pulse and put $\alpha_{r}=+2$ for the leading pulse. This condition gives us high contrast fringes with the same measured curvature shown in fig. \ref{fig:evol}(b).\\
  The temporal delays of the two gaussian pulses in fig. \ref{fig:evol_num}(b) where chosen so as to match the experimental fringe frequency, so that $\tau=-220$  and $\tau=220$ fs with respect to a central X-wave. As already discussed, we are to expect pulse splitting or dynamical replenishment for these higher input energies. The actual position of the X component does not induce any relevant changes in the AE and we are not able to precisely determine the temporal location of the CE. This may be explained by noting that the total power contained in the experimental spectrum related to CE part (after the monocromator input slit) is always at least a factor ten smaller than that due to AE, so that interference between the two contributions is rather weak (see for example the very low contrast interference fringes in figs. \ref{fig:evol}(a) and \ref{fig:evol_num}(a)). The total delay between the leading and trailing pulses (440 fs) is surprisingly high if compared to the 200 fs input pulse duration. Obviously GVD is playing a major role and is indeed dominating the temporal profile evolution. \\ 
  Figure \ref{fig:evol_num}(c) was obtained with three such gaussian pulses with temporal delays $\tau=-400$, $-240$, $+400$ and $\alpha=+2$, $+2$ and $-2$ respectively. Once again we were not able to precisely position the X-wave as negligible variations were observed. The agreement with the experimental data (Fig. \ref{fig:evol}(c)) is very good and the model is in agreement with other reported findings relative to the formation of multiple peak re-formation for high enough input energies \cite{mlejnek:1999, liu:2003}. Furthermore, the temporal delay between the pulses has increased to 800 fs, i.e. four times the input pulse duration and indicates that for higher powers the interplay between SPM, SC generation and GVD is further enhanced, as expected. This delay is extremely large if compared to the input pulse duration however a further indication that this result is correct is given by the multiple-shot autocorrelation trace shown in fig. \ref{fig:autocor} obtained by imaging the output crystal facet onto the autocorrelation nonlinear crystal. The five peaks in the trace show that with a similar average pulse input energy we have multiple pulse splitting with the formation of three daughter pulses and a temporal separation of 600 fs between the leading and trailing pulses.\\
  So far we have purposely neglected temporal chirp ($\alpha_{t}$) which is expected to be at least as important as the spatial chirp. However AE does not seem to be sensitive to this parameter. Indeed, variations of $\alpha_{t}$ did not produce a significant change in the numerical AE fringe pattern but rather only a local reduction of contrast, the position of which depends on the sign and value of $\alpha_{t}$ in the various linear components with which the filament was modeled.
 We note though, that the experimental spectra also show a modulation in the CE in the form of a multiple ``X-arm'' splitting. In fact, by slowly increasing the pulse input energy from below to above threshold we never experimentally observed the formation of a single X pattern - CE emission in our experimental setup always seems to appear in the form of multiple X-like patterns.  This sheds some light on the nature of the CE in the sense that it may be explained by super-imposing multiple X pulses so that the number of arms corresponds directly to the number of X pulses and each X must be spectrally shifted with respect to the others. This is physically feasible in the presence of a single input pulse that undergoes sever temporal chirping and then splits into multiple ``X-pulses'' (i.e. daughter pulses that gives rise to CE) so that each of these is centered on a different carrier frequency. The values for the spectral shift observed in our spectra vary from 0.05 to 0.2 fs$^{-1}$ with the higher values observed at higher input energies.\\
 
 \section{Conclusions}
In conclusion we have shown the possibility of recovering detailed information regarding space-time coupled phenomena from angular-spectral characterization. The nature of the numerical model used to interpret the experimental measurements, based on a linear combination of stationary states rather than on a full nonlinear evolution simulation, allows only a qualitative guess at the parameters involved. Nevertheless the simplicity of this approach is very appealing and indeed, by using a combination of linear states it is possible to understand if the pulse has changed its spatio-temporal energy distribution without actually experiencing splitting or if it has gone through a single or even multiple pulse reformation. The details of the spectra also reveal that the spatio-temporal coupling manifests itself in a strong spatial chirp of the filament. If the far-field measurements were to be completed with a complete phase characterization we would of course be able to reconstruct the full space-time profile of the pulse through a Fourier transform - the lack of this information is reflected in the difficulty in retrieving more precise information for example of the exact temporal delay of the CE with respect to the AE sources. Furthermore, the spatial chirp parameter may be determined precisely with the combination of an imaging spectrometer and a spatial shearing interferometer \cite{dorrer:2002} and shall be considered in future measurements. We further underline that whereas the ``X'' part of the spectra are numerically well reproduced in the frame of our simple linear model in both the Stokes and anti-Stokes regions, this is not true for the AE spectral components. The anti-Stokes AE is rather confused and the (not always visible) fringe pattern seems to show a fast modulation frequency that is not compatible with the parameters that fit the Stokes region. This could indicate that the shock waves that generate the blue and red shifts in the spectrum are spatially distinct and have different temporal delays.\\
Notwithstanding these shortcomings, the aim of this work is to underline the importance of angular spectral measurements that, together with other methods such as FROG or 3D-mapping, may give a complete and exhaustive characterization of nonlinear optical-wave collapse phenomena.\\
$^{**}$ Present address, Institut de Ci\`encies Fot\`oniques, c/Jordi Girona 29, NEXUS II, E-08034 Barcelona, Spain.\\
$^{***}$ Present address, Laboratoire Kastler Brossel, Universit\'e P. et M. Curie, Case74, 4, place Jussieu, F75252 Paris, France

%\begin{acknowledgments}

%\end{acknowledgments}

\newpage
%\bibliography{filamentation}

\begin{thebibliography}{10}
\newcommand{\enquote}[1]{``#1''}
\expandafter\ifx\csname url\endcsname\relax
  \def\url#1{{#1}}\fi
\expandafter\ifx\csname urlprefix\endcsname\relax\def\urlprefix{}\fi

\bibitem{ask:1962}
G.~A.~Askarjan, \enquote{Interaction of laser radiation with vibrating surfaces,} JETP, {\bf 42}, 1672 (1962).

\bibitem{talanof:1964}
W.~Talanov, Izv. Vysshikh Uchebn. Zavedenii, Radiofizika {\bf 7} (1964).

\bibitem{chiao:1964}
R.Y.Chiao, E.~Garmire, and C.H.Townes, \enquote{Self-trapping of optical
  beams,} Phys. Rev. Lett. {\bf 13}, 479--482 (1964).

\bibitem{hercher:1964}
M.~Hercher, J. Opt. Soc. Am. {\bf 54}, 563 (1964).

\bibitem{braun:1995}
A.~Braun, G.~Korn, X.~Liu, D.~Du, J.~Squier, and G.Mourou,
  \enquote{Self-channeling of high-paek-power femtosecond laser pulses in air,}
  Opt. Lett. {\bf 20}, 73--75 (1995).

\bibitem{wilson:1997}
K.~Wilson and V.~Yakovlev, \enquote{Ultrafast rainbow: tunable ultrashort
  pulses from a solid-state kilohertz system,} J. Opt. Soc. B {\bf 14},
  444--448 (1997).

\bibitem{alfano:1989}
R.~Alfano, {\em The Supercontinuum Laser Source\/} (Springer-Verlag, New York,
  1989).

\bibitem{rai:2000}
P.~Rairoux, H.~Schillinger, S.~Niedermeier, M.~Rodriguez, F.~Ronneberger,
  R.~Sauerbrey, B.Stein, D.~Waite, C.~W. nd~H.~Wille, L.~W\"oste, and C.Ziener,
  \enquote{Remote sensing of the atmosphere using ultrashort laser pulses,}
  Appl. Phys. B {\bf 71}, 573--580 (2000).

\bibitem{noz:1990}
P.~Nozieres and D.~Pines, {\em The Theory of Quantum Liquids, vol. II\/}
  (Addison-Wesley, Redwood City, 1990).

\bibitem{strickland:1994}
D.Strickland and P.~Corkum, \enquote{Resistance of short pulses to
  self-focusing,} J.Opt.Soc.Am.B {\bf 11}, 492--497 (1994).

\bibitem{luther:1994}
G.~Luther, A.~Newell, J.~Moloney, and E.~Wright, \enquote{Short pulse conical
  emission and spectral broadening in normally dispersive media,} Opt. Lett.
  {\bf 19}, 789--791 (1994).

\bibitem{gaeta:2000}
A.~Gaeta, \enquote{Catastrophic collapse of ultrashort pulses,} Phys. Rev.
  Lett. {\bf 84}, 3582--3585 (2000).
 

\bibitem{ranka:1998}
J.~Ranka and A.~Gaeta, \enquote{Breakdown of the slowly varying envelope
  approximation in the self-focusing of ultra-short pulses,} Opt. Lett. {\bf
  23}, 534--536 (1998).

\bibitem{trillo:2002}
S.~Trillo, C.~Conti, P.~D. Trapani, O.~Jedrkiewicz, J.~Trull, G.~Valiulis, and
  G.~Bellanca, \enquote{Coloured conical emission via second-harmonic
  generation,} Opt. Lett. {\bf 27}, 1451--1453 (2002).

\bibitem{valiulis:2001}
G.~Valiulis, J.~Kilius, O.~Jedrkiewicz, A.~Bramati, S.~Minardi, C.~Conti, S.~Trillo, A.~Piskarskas, and  P.~Di~Trapani, \enquote{Space-time nonlinear compression and three-dimensional complex trapping in normal dispersion,"} in OSA Trends in Optics and Photonics (TOPS) {\bf 57}, Quantum Electronics and Laser Science Conference (QELS 2001), Technical Digest, Post conference Edition (Optical Society of America, Washington DC, 2001), pp. QPD10 -1-2.

\bibitem{ditrapani:2003}
P.~Di~Trapani, G.~Valiulis, A.~Piskarskas, O.~Jedrkiewicz, J.~Trull, C.~Conti,
  and S.~Trillo, \enquote{Spontaneously generated X-shaped light bullets,}
  Phys. Rev. Lett. {\bf 91}, 093\,904--1 (2003).

\bibitem{conti:2003}
C.~Conti, S.~Trillo, P.~Di~Trapani, G.~Valiulis, A.~Piskarskas, O.~Jedrkiewicz,
  and J.~Trull, \enquote{Nonlinear electromagnetic X waves,} Phys. Rev. Lett.
  {\bf 90}, 170\,406--1 (2003).

\bibitem{ottavia:2003}
O.~Jedrkiewicz, J.~Trull, G.~Valiulis, A.~Piskarskas, C.~Conti, S.~Trillo, and
  P.~Di~Trapani, \enquote{Nonlinear X wavesin second harmonic generation:
  esperimental results,} Phys. Rev. E {\bf 68}, 026\,610--1 (2003).

\bibitem{salerno:2004}
D.Salerno, O.~Jedrkiewicz, P.~Di~Trapani, J.~Trull, and G.~Valiulis,
  \enquote{Impact of dimensionality on noise-seeded modulational instability,}
  in {\em Nonlinear guided waves and their applications\/} (2004), pp. WA--7.

\bibitem{minardi:2003}
S.~Minardi, J.~Yu, G.~Blasi, A.~Varanavi\v{c}ius, G.~Valiulis,
  A.~Ber\v{z}anskis, A.~Piskarskas, and P.~Di~Trapani, \enquote{Red solitons:
  evidence of spatiotemporal instability in {$\chi^{(2)}$} spatial soliton
  dynamics,} Phys. Rev. Lett. {\bf 91}, 12\,390--1 (2003).

\bibitem{kolesik:arxiv2003}
M.~Kolesik, E.~Wright, and J.~Moloney, \enquote{Dynamic Nonlinear X-waves for
  Femtosecond Pulse Propagation in Water,} Arxiv:Phys/0311021v1  (2003).

%\bibitem{porras:2004}
%M.~Porras, A.~Parola, A.~Dubietis, and P.~Di Trapani, \enquote{Stationary
%  conical waves supported by nonlinear absorption,} submitted to Phys. Rev.
%  Lett.  (2004).
  
  \bibitem{dub:2004}
A.~Dubietis, E.~Gaizauskas, G.~Tamosauskas, and P.~Di~Trapani, \enquote{Light
filaments without self-channeling,} accepted for publication in Phys. Rev.
  Lett.  (2004).

\bibitem{diddams:1998}
S.~Diddams, H.~Eaton, A.~Zozulya, and T.~Clement, \enquote{Amplitude and phase
  measurements of femtosecond pulse splitting in nonlinear dispersive media,}
  Opt. Lett. {\bf 23}, 379--381 (1998).

\bibitem{bernstein:2003}
A.~Bernstein, J.~Diels, T.~Luk, T.~Nelson, A.~McPherson, and S.~Cameron,
  \enquote{Time resolved measurements of self focusing pulses in air,} Opt.
  Lett. {\bf 28}, 2354--2356 (2003).

\bibitem{gallmann:2001}
L.~Gallmann, G.~Steinmeyer, D.~Sutter, T.~Rupp, C.~Iaconis, I.~Walmsley, and
  U.~Keller, \enquote{Spatially resolved amplitude and phase characterization
  of femtosecond optical pulses,} Opt. Lett. {\bf 26}, 96--98 (2001).

\bibitem{dorrer2:2002}
C.~Dorrer, E.~Kosik, and I.~Walmsley, \enquote{Direct space-time
  characterization of the electric fields of ultrashort optical pulses,} Opt.
  Lett. {\bf 27}, 548--550 (2002).

\bibitem{potenza:2004}
M.~Potenza, S.~Minardi, J.~Trull, G.~Blasi, D.~Salerno, A.~Varanavi\v{c}ius,
  A.~Piskarskas, and P.~Di~Trapani, \enquote{Three dimensional imaging of short
  pulses,} Opt. Commun. {\bf 229}, 381--390 (2004).

\bibitem{trull:2004}
J.~Trull, O.~Jedrkiewicz, P.~D. Trapani, A.~Matijosius, A.~Varanavi\v{c}ius,
  G.~Valiulis, R.~Danielius, E.~Kucinskas, and A.~Piskarskas,
  \enquote{Spatiotemporal three-dimensional mapping of nonlinear X-waves,}
  Phys. Rev. E {\bf 69}, 026\,607--1 (2004).

\bibitem{mat:2004}
A.~Matijosius, J.~Trull, P.~Di~Trapani, A.~Dubietis, R.~Piskarskas,
  A.~Varanavi\v{c}ius, and A.~Piskarskas, \enquote{Nonlinear space-time dynamics
  of ultrashort wave packets in water,} Opt. Lett. {\bf 29}, 1123--1125 (2004).

\bibitem{mikalauskas:2002}
D.~Mikalauskas, A.~Dubietis, and R.~Danielius, \enquote{Observation of light
  filaments induced in air by visible picosecond laser pulses,} Appl. Phys. B
  {\bf 75}, 899--902 (2002).

\bibitem{dub:2003}
A.~Dubietis, G.~Tamo\v{s}auskas, I.~Diomin, and A.~Varanavi\v{c}ius,
  \enquote{Self-guided propagation of femtosecond light pulses in water,} Opt.
  Lett. {\bf 28}, 1269--1271 (2003).

\bibitem{modos}
M.~Porras and P.~D. Trapani, \enquote{Localized and stationary light wave modes
  in dispersive media,} Los Alamos Arxive physics/0309084  (2003).

\bibitem{marburger:1975}
J.~Marburger, \enquote{Self-focusing: theory,} Progr. Quant, Electron. {\bf 4},
  35--110 (1975).

\bibitem{chernev:1992}
P.~Chernev and V.~Petrov, \enquote{Self-focusing of light pulses in th
  presence of normal group-velocity dispersion,} Opt. Lett. {\bf 17}, 172--174
  (1992).

\bibitem{luther2:1994}
G.~Luther, J.~Moloney, A.~Newell, and E.~Wright, \enquote{Self-focusing
  threshold in normally dispersive media,} Opt. Lett. {\bf 19}, 862--864
  (1994).

\bibitem{brodeur:1999}
A.~Brodeur and S.~Chin, \enquote{Ultrafast white light continuum generation and
  self-focusing in transarent condensed media,} J.Opt.Soc.Am.B {\bf 16},
  637--650 (1999).

\bibitem{zozulya:1999}
A.~Zozulya and S.A.Diddams, \enquote{Dynamics of self-focused femtosecond laser
  pulses in the near and far fields,} Opt. Expr. {\bf 4}, 336--343 (1999).

\bibitem{rothen:1992}
J.~Rothenberg, \enquote{Space-time focusing: breakdown of the slowly varying
  envelope approximation in the self-focusing of femtosecond pulses,} Opt.
  Lett. {\bf 17}, 1340--1342 (1992).

\bibitem{mlejnek:1998}
M.~Mlejnek, E.~Wright, and J.~Moloney, \enquote{Dynamic spatial replenishment
  of femtosecond pulses propagating in air,} Opt. Lett. {\bf 23}, 382--384
  (1998).

\bibitem{mlejnek:1999}
M.~Mlejnek, E.~Wright, and J.~Moloney, \enquote{Power dependence of dynamic
  spatial replenishment of femtosecond pulses propagating in air,} Opt. Expr.
  {\bf 4}, 223--228 (1999).

\bibitem{kosareva:1997}
O.~Kosareva, V.~Kandidov, A.~Brodeur, C.~Chen, and S.~Chin, \enquote{Conical
  emission from laser-plasma interactions in the filamentation of powerful
  ultrashort laser pulses in air,} Opt. Lett. {\bf 22}, 1332--1334 (1997).

\bibitem{tzortzakis:2001}
S.~Tzortzakis, L.~Sudrie, M.~Franco, B.Prade, A.~Mysyrowicz, A.~Couairon, and
  L.~Berg\'e, \enquote{Self-guided propagation of ultrashort IR laser pulses
  in fused silica,} Phys. Rev. Lett. {\bf 87}, 3902--1 (2001).

\bibitem{bloembergen:1977}
W.~Smith, P.~Liu, and N.~Bloembergen, \enquote{Superbroadening in {$H_{2}O$}
  and {$D_{2}O$} by self focused picosecond pulses from a {YAlG:Nd} laser,}
  Phys. Rev. A {\bf 15}, 2396--2403 (1977).

\bibitem{nibbering:1996}
E.~Nibbering, P.~Curley, G.~Grillon, B.~Prade, M.~Franco, F.Salin, and
  A.Mysyrowicz, \enquote{Conical emission from self guided femtosecond pulses
  in air,} Opt. Lett. {\bf 21}, 62--64 (1996).

\bibitem{liu:2003b}
W.~Liu, O.~Kosareva, I.~Golubtsov, A.~Iwasaki, A.~Becker, V.~Kandidov, and
  S.~Chin, \enquote{Femtosecond laser pulse filamentation versus optical
  breakdown in {$H_{2}O$},} Appl.Phys. B {\bf 76}, 215--229 (2003).

\bibitem{cundiff:1996}
S.~Cundiff, W.~Knox, E.~Ippen, and H.~Haus, \enquote{Frequency dependent mode
  size in broadband Kerr-lens mode locking,} Opt. Lett. {\bf 21}, 662--664
  (1996).

\bibitem{polyakov:2001}
S.~Polyakov, F.~Yoshino, G.~Stegeman, \enquote{Interplay between self-focusing and high-order multiphoton absorption,} J. Opt. Soc. Am. B {\bf 18}, 1891--1895
  (2001).

\bibitem{akhmanov}
S.~Akhmanov, V.~Vysloukh, and A.~Chirkin, {\em Optics of Femtosecond Laser
  Pulses\/} (American Institute of Physics, New York, 1992).

\bibitem{liu:2003}
W.~Liu, S.~Chin, O.~Kosareva, I.S.Golubtsov, and V.~Kandidov, \enquote{Multiple
  refocusing of a femtosecond laser pulse in a dispersive liquid (methanol),}
  Opt. Commun. {\bf 225}, 193--209 (2003).

\bibitem{dorrer:2002}
C.~Dorrer and I.~Walmsley, \enquote{Simple linear tecnique for the measurement
  of space-time coupling in ultrashort optical pulses,} Opt. Lett. {\bf 27},
  1947--1949 (2002).

\end{thebibliography}

\end{document}